\documentclass[preprint,3p,12pt]{elsarticle}
\usepackage{lineno,hyperref}
\usepackage{mathrsfs}
\usepackage{amsmath}
\biboptions{sort&compress}
\usepackage{verbatim} 
\usepackage{epsfig}
\usepackage{amssymb} 
\usepackage{mathtools}
\usepackage{lineno}
\usepackage{tikz}     
\usepackage{etoolbox}
\usepackage{ulem} 
\usepackage{textcomp}
\usepackage{nicematrix}
\usepackage{silence}
\usepackage{hyperref}
\usepackage{float}
\usepackage{subcaption}
\usepackage{graphicx}
\usepackage{placeins}

\makeatletter \def\ps@pprintTitle{   \let\@oddhead\@empty   \let\@evenhead\@empty   \let\@oddfoot\@empty   \let\@evenfoot\@empty } \makeatother

\begin{document}

\newcommand*{\hwplotB}{\raisebox{3pt}{\tikz{\draw[red,dashed,line 
width=3.2pt](0,0) -- 
(5mm,0);}}}

\newrobustcmd*{\mydiamond}[1]{\tikz{\filldraw[black,fill=#1] (0.05,0) -- 
(0.2cm,0.2cm) -- (0.35cm,0) -- (0.2cm,-0.2cm) -- (0.05,0);}}

\newrobustcmd*{\mytriangleright}[1]{\tikz{\filldraw[black,fill=#1] (0,0.2cm) -- 
(0.3cm,0) -- (0,-0.2cm) -- (0,0.2cm);}}

\newrobustcmd*{\mytriangleup}[1]{\tikz{\filldraw[black,fill=#1] (0,0.3cm) 
-- (0.2cm,0) -- (-0.2cm,0) -- (0,0.3cm);}}

\newrobustcmd*{\mytriangleleft}[1]{\tikz{\filldraw[black,fill=#1] (0,0.2cm) -- 
(-0.3cm,0) -- (0,-0.2cm) -- (0,0.2cm);}}
\definecolor{Blue}{cmyk}{1.,1.,0,0} 

\begin{frontmatter}

\title{Modeling Epidemics on Multiplex Networks:\\Epidemic Threshold and Basic Reproduction Number}

\address[IB]{Instituto Balseiro, Universidad Nacional de Cuyo, Argentina.}
\address[CNEA]{Física Estadística e Interdisciplinaria, Centro Atómico Bariloche, Comisión Nacional de Energía Atómica. }
\address[CONICET]{Consejo Nacional de Investigaciones Científicas y Técnicas (CONICET), Argentina.}
\address[Tandil]{Universidad Nacional del Centro de la Provincia de Buenos Aires, Facultad de Cs. Exactas, Instituto Multidisciplinario Sobre Ecosistemas Y Desarrollo Sustentable, Tandil, Argentina.}
\address[Tandil2]{Universidad Nacional del Centro de la Provincia de Buenos Aires, Facultad de Cs. Exactas, Instituto PLADEMA, Tandil, Argentina.}

\author[IB,CNEA,CONICET]{E.A.~Rozan}
\ead{rozan.eric@gmail.com}

\author[CONICET,Tandil,Tandil2]{M.I.~Simoy}
\ead{ignacio.simoy@gmail.com}

\author[CNEA,CONICET]{S.~Bouzat}
\ead{bouzat@cab.cnea.gov.ar}

\author[IB,CNEA,CONICET]{M.N.~Kuperman}
\ead{marcelo.kuperman@ib.edu.ar}

\begin{abstract}
\nolinebreak[0] Accurate epidemic forecasting requires models that account for the layered and heterogeneous nature of real social interactions. The basic reproduction number $\mathcal R_0$, as calculated from models that assume homogeneous mixing or single-layer contact structures, has limited applicability to complex social systems. Here, we derive an expression for $\mathcal R_0$ in the context of multiplex networks, enabling the analysis of disease transmission across multiple social layers.

We adapt the Degree-Based Mean-Field (DBMF) SIR model for single-layer complex networks to the multiplex setting, where each layer is characterized by its own degree distribution and infection rate. Using the Next Generation Matrix method, we derive an analytical expression for the basic reproduction number $\mathcal R_0$. Numerical integration of the multiplex DBMF equations shows that $\mathcal R_0=1$ marks the epidemic threshold and governs the behavior of key outbreak indicators as expected. In addition to the exact expression for $\mathcal R_0$, we introduce an approximation, denoted by $\tau$, which is simpler to compute and admits a more transparent interpretation in terms of the epidemiological and topological parameters of the system.

Stochastic agent-based simulations support these findings, demonstrating a direct correspondence between $\tau$ and the average number of secondary infections generated during the early stages of an outbreak, consistent with the epidemiological interpretation of $\mathcal R_0$. This work provides a robust generalization of $\mathcal R_0$ for layered contact structures, offering a more realistic basis for epidemic forecasting and the design of intervention strategies.
\end{abstract}


\end{frontmatter}

\setlength{\parskip}{12pt}
\section{Introduction} \label{section:intro}

In mathematical epidemiology, the basic reproduction number $\mathcal R_0$ is a fundamental epidemiological indicator that quantifies the average number of secondary infections generated by a single infected individual in a fully susceptible population. It also serves as a critical threshold parameter: when $\mathcal R_0 > 1$, an epidemic is likely to occur, whereas if $\mathcal R_0 < 1$, the disease tends to die out. Furthermore, this quantity determines relevant indicators such as the peak of infections and the epidemic size~\cite{kerm1,murray}. This measure guides public health responses by informing vaccination coverage targets and control strategies, and it provides a comparative framework for assessing the transmission potential of different pathogens. However, $\mathcal R_0$ is often estimated from models that do not account for population heterogeneity, temporal changes in susceptibility, behavioral adaptations, or pathogen evolution, all of which can significantly influence disease spread~\cite{R0_heterogeneo,may_HIV}. Meanwhile, advances in mathematical modeling, including the use of next-generation matrices and statistical approaches, enable more robust estimation of reproduction numbers in heterogeneous and complex populations~\cite{next_gen}.

Since the beginning of the century, studying epidemic processes on complex networks has revealed emerging patterns associated with the heterogeneity of the social structure \cite{review_networks,kuperman_2013}. The core idea of network-based disease spread models is that transmission occurs only along existing network edges, which represent a contact between individuals. 

Agent-based models can explicitly consider a contact structure, with results varying depending on the topology of the considered network \cite{networks_and_epidemic,kuperman,zanette,zanette2,kara,multi_agent,benitez}. The connectivity patterns can also be incorporated into mean-field differential equations models by considering key aspects of the social structure, like the degree distribution of the underlying network~\cite{moreno2002, dynamical_patterns, inmunization_pastor_satorras, velocity_spread,rozan_2023}.

More recently, multiplex networks have been used to capture more intricate behaviors~\cite{singapore,liuq,overlapped_multiplex_networks}. Each layer of the network represents a different type of social interaction between agents, allowing for a more detailed characterization. For instance, this approach enables the separate modeling of contacts occurring in schools, workplaces, public spaces, and households. While each layer has its own distinct dynamics, the interconnection of these networks shapes the overall evolution of the epidemic outbreak. This layered representation reflects more realistically the fact that individuals interact differently across various social contexts, which significantly influences disease transmission dynamics. This approach can capture super-spreading events, and heterogeneity in individual infectiousness.

In this sense, it is clear that estimating $\mathcal R_0$ through models based on homogeneous mixing can be inadequate or inaccurate.  In this paper, we aim to calculate the basic reproduction number $\mathcal R_0$ within the framework of multiplex networks. By integrating the structural complexity of multiplex networks into epidemic modeling, we seek to provide a more realistic and context-sensitive estimation of $\mathcal R_0$, thereby improving our understanding of disease spread and informing more effective intervention strategies.

This paper is arranged as follows. In the next Section, we introduce the theoretical background about epidemiological indicators on the SIR model using ordinary differential equations and the degree-based mean-field model (DBMF) of epidemics in the single-layered network model presented in \cite{moreno2002,velocity_spread}. In Section~\ref{section:multiplex_model}, we adapt the DBMF model to include the degree distributions of the layers that constitute a multiplex network parting from the single layered model. In addition, we use the standard Next Generation Matrix method to derive an analytical expression for $\mathcal R_0$ , which accounts for the multiple layers of the network. Section~\ref{section:results_ode} shows results coming from numerically integrating the differential equations of the model, while Section~\ref{section:results_agents} focuses on results coming from stochastic agent-based simulations. Finally, we present the discussion of the results and conclusions in Section~\ref{section:conclusion}.

\section{Theoretical Background}\label{section:background}

\begin{figure}[ht]\centering
 \includegraphics[width=0.5\linewidth]{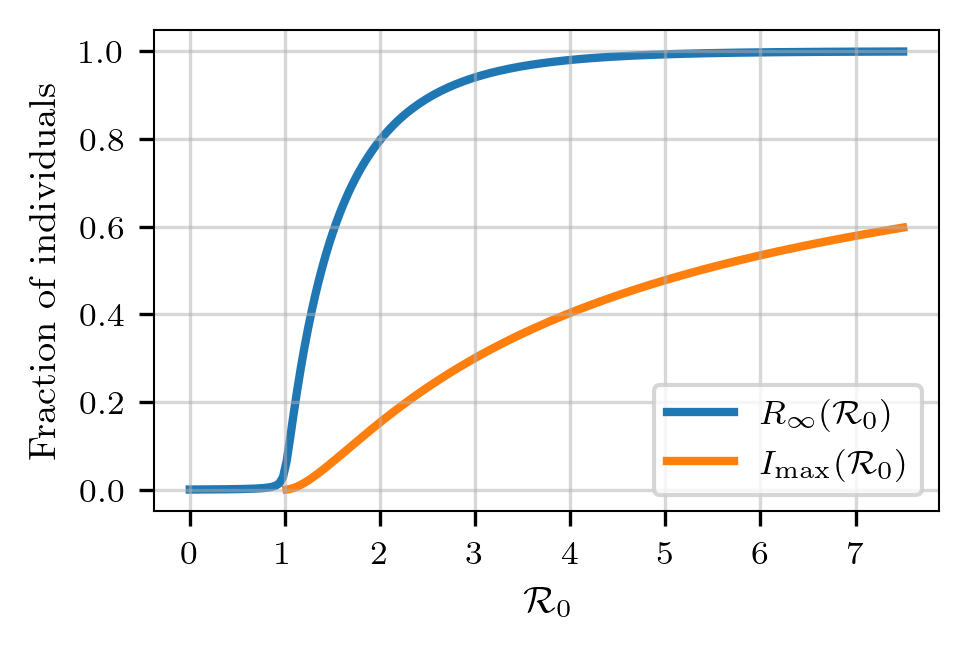}
 \caption{Maximum infection peak $\displaystyle I_\mathrm{max}= \max_t \;I_\mathrm{tot}(t)$ and final epidemic size $R_\infty= R_\mathrm{tot}(t\to\infty)$ attained in the standard SIR model. Their analytical expressions are $I_\mathrm{max} = 1-\frac{1+\ln  \left(S(0) \mathcal R_0\right)}{\mathcal R_0}$ and ${R_\infty = 1 + \frac{W\left( - S(0) \mathcal R_0 e^{-\mathcal R_0}\right)}{\mathcal R_0}} $ respectively~\cite{murray}, where $W$ stands for the Lambert $W$ function. }
 \label{ind_SIR}
\end{figure}

The basic reproduction number $\mathcal R_0$ is most straightforwardly derived in the standard SIR model, described by the equations
\refstepcounter{equation} 
\begin{alignat}{2}
\frac{\text d S}{\text d t} =&-\beta SI\tag{\theequation.a} \\
\frac{\text d I}{\text d t} =& \beta SI- \gamma  I \tag{\theequation.b}\\
\frac{\text d R}{\text d t} =&\quad \gamma I \tag{\theequation.c}
\end{alignat}
\noindent where $S$, $I$, and $R$ are the fraction of individuals in the Susceptibles, Infectious and Recovered compartments, respectively, $\beta$ is the infection rate and $\gamma$ is the recovery rate. In this model, $\mathcal R_0=\beta/\gamma$~\cite{kerm1,murray}. The maximum infection peak $\displaystyle I_\mathrm{max}= \max_t \;I_\mathrm{tot}(t)$ and the final epidemic size $R_\infty= R_\mathrm{tot}(t\to\infty)$ are completely determined by $\mathcal R_0$ as shown in Fig.~\ref{ind_SIR}~\cite{murray}.

\subsection{DBMF model for heterogeneous populations in complex networks}

The starting point of the present work is the Degree-Based Mean-Field (DBMF) SIR model~\cite{moreno2002,velocity_spread}, in which the population is represented as a complex network and the degree distribution plays a key role. The fraction of individuals that have $k$ daily contacts is $P(k)$, where $k$ attains values between $1$ and $K$ (the maximum degree attained by any individual).

The Susceptible, Infectious, and Recovered/Removed epidemiological compartments are divided into ${K}$ sub-compartments each. For example, the S compartment is divided into S$_1$,S$_2$,...,S$_{{K}}$, where $S_k$ is the fraction of individuals with $k$ daily contacts that are susceptible. Analogously, $I_k$, and $R_k$ are the fractions of individuals with degree $k$ in the respective epidemiological state. This implies $S_k+I_k+R_k=1\:\forall k$. 

The total fraction of individuals in the $x$ compartment (where $x$ stands for any of $S$, $I$ or $R$), is computed as follows:
\begin{equation}
x_\text{tot}=\sum_{k=1}^{{K}}\:P(k)\:x_k 
 \label{rho_tot}
\end{equation}

The equations for the time evolution of the $3{K}$ sub-compartments of the model are

\refstepcounter{equation} 
\addtocounter{equation}{-1} 
\begin{subequations}\label{network_model}
\begin{alignat}{2}
\frac{\text d S_k}{\text d t} =&\quad -\beta kS_k\theta_k \label{Sk} \tag{\theequation.a} \\
\frac{\text d I_k}{\text d t} =&\quad \beta kS_k\theta_k- \gamma  I_k \tag{\theequation.b}\\
\frac{\text d R_k}{\text d t} =&\quad \gamma I_k \label{Rk}\tag{\theequation.c}
\end{alignat}
\end{subequations}

\noindent where $\beta$ is the infection rate, $\gamma$ is the recovery rate (\textit{i.e.}, inverse of the mean recovery time), and $\theta_k$ is the density of infected neighbors of an individual with degree $k$. For uncorrelated networks, $\theta_k$ does not depend on the degree $k$, and it can be expressed as follows~\cite{velocity_spread}:
\begin{equation}
     \theta(t) = \sum_{k'=1}^K \frac{(k'-1) \:P(k')\: I_{k'}(t)}{\langle k \rangle} \label{theta}
\end{equation}
where $\langle k\rangle$ is the mean degree $\displaystyle \sum_{k=1}^K kP(k)$. The initial conditions are $I_k = I_0\ll 1$, ${S_k=1-I_0}$, and $R_k=0\:\forall k$.

\subsection{Basic reproduction Number in the Single-Layered Model}

The basic basic reproduction number $\mathcal R_0$ of the model can be obtained using the Next Generation Matrix (NGM) method~\cite{next_gen}. Firstly, we order the compartments so that the vector of dynamical variables is $ X = (I_1,\cdots I_K, S_1,\cdots,S_K,R_1,\cdots,R_K) $. Then, we write the system of equations~(\ref{network_model}) as $\dot X = \mathscr F-\mathscr V$ with

\begin{equation}
    \mathscr F_i =\left\{
    \begin{array}{cc}
        \beta iS_i\theta & i\leq K \\
         0 & K <i\leq3K
    \end{array} \right.\quad\quad \mathscr V_i = \left\{
    \begin{array}{cc}
        \gamma I_i & i\leq K \\
         \mathscr F_{i-K} &  \ K<i\leq 2K\\
         -\gamma I_{i-2K }& 2K <i\leq3K
    \end{array} \right.
\end{equation}

Following the NGM method, we proceed to compute the matrices $F$ and $V$ defined as
\begin{equation}
F_{i,j}=\frac{\partial \mathscr F_i}{\partial I_j} (X=X^*)= \beta i \frac{\partial \theta}{\partial I_j} = \beta  i  \sum_{k'=1}^K \frac{(k'-1)  P(k')}{\langle k \rangle} \delta_{j,k'} =  \beta \  i\  \frac{(j-1)  P(j)}{\langle k \rangle}
\end{equation}

and 

\begin{equation}
    V_{i,j} = \frac{\partial\mathscr V_i}{\partial I_j} (X=X^*)= \gamma\: \delta_{i,j} \Rightarrow V^{-1}_{i,j } = \frac1\gamma\ \delta_{i,j}\quad \mathrm{for}\ i,j\leq K
\end{equation}
\noindent where $X^*$ represents the disease free equilibrium, with $S_k^*=1$, $I_k^*=R^*_k=0\:\forall k$, and $\delta_{i,j}$ being the Kronecker delta. According to the NGM method, $\mathcal{R}_0=\rho\left(FV^{-1}\right)$, meaning the spectral radius of the matrix $M=FV^{-1}$, given by
\begin{equation}
    M_{i,j} = \left(FV^{-1}\right)_{i,j} = \frac{\beta}{\gamma \, \langle k \rangle}  \;i\,(j-1)\,P(j)\: 
\end{equation}

Notice that, for simplicity, we can define $\alpha =  \frac{\beta}{\gamma \, \langle k \rangle} $, the vector $\vec v$ given by $v_i=i$ and the vector $\vec u$ given by $u_j=(j-1)P(j)$ to rewrite $M$ as 

\begin{equation}
    M_{i,j} = \alpha\ v_i\,u_j \quad \Rightarrow\quad  M= \alpha \ \vec v\, \vec u^T
\end{equation}

By expressing $M$ as the outer product of two vectors, it becomes evident that its rank is $1$. Consequently, its only non-zero eigenvalue is equal to its trace:

\begin{equation}
  \mathcal R_0 = \rho(M)=\mathrm{Tr}(M) = \sum_{k=1}^K \alpha\ v_k\, u_k =  \alpha\ \sum_{k=1}^K k(k-1)P(k)=\frac{\beta}{\gamma} \frac{\langle k^2 \rangle-\langle k \rangle}{\langle k \rangle} \label{R0_monoplex}
\end{equation}

Which is consistent with the literature on single-layered networks~\cite{moreno2002,brauer}.

\section{DBMF Multiplex Model}\label{section:multiplex_model}
The objective of this section is to extend the model discussed above to multiplex networks, to account for the same infection being able to be transmitted through different contact structures. In other words, although there is only one infection being propagated through the population, the topology and the infection rate of each layer may be different. This work will focus on the case of two-layered networks. 

To begin with, the degree distribution should now account for the degree of the nodes in each layer. The joint degree distribution $P(k_1,k_2)$, with $k_1=1,2,...,K_1$ and $k_2= 1,2,...,K_2$, represents the fraction of individuals that have $k_1$ contacts in the layer $1$ and $k_2$ contacts in the layer $2$. Throughout this work, we assume that the layers are statistically independent, so that $P(k_1,k_2)=P_1(k_1)P_2(k_2)$, where $P_1(k_1)$ and $P_2(k_2)$ denote the degree distributions of layers $1$ and $2$, respectively. This assumption is motivated by the nature of the social contexts represented by the layers. For example, the number of household contacts of an individual does not necessarily determine the number of contacts they have in professional, educational, or recreational environments. Therefore, statistical independence provides a reasonable first approximation for the systems considered here. 

Following a similar approach to that of the previous model, each epidemiological compartment is divided into $K_1\times K_2$ sub-compartments. If $x$ represents any of the compartments (S, I or R), then $x_{k_1,k_2}$ represents the fraction of the individuals that have $k_1$ and $k_2$ contacts in layer $1$ and $2$, respectively, and is in the $x$ compartment. Thus, $S_{k_1,k_2}+I_{k_1,k_2}+R_{k_1,k_2}=1\:\forall k_1,k_2$. 
Marginalization can be done to obtain, for example, the fraction of the individuals in any compartment $x$ that have $k_1$ contacts in layer $1$: $\displaystyle x_{k_1}=\sum_{k_2}   P_2(k_2)\,x_{k_1,k_2}$. The expression for the total density of individuals in the $x$ compartment is given by
\begin{equation}
x_\mathrm{tot} = \sum_{k_1=1}^{K_1}\sum_{k_2=1}^{K_2} \;P(k_1,k_2)\;x_{k_1,k_2} \label{rho_tot_multiplex}
\end{equation}

\refstepcounter{equation}
\addtocounter{equation}{-1} 

The differential equations for the evolution of the sub-compartments are as follows:
\begin{subequations}\label{multiplex_model}
\begin{alignat}{2}
\frac{\text d S_{k_1,k_2} }{\text d t} =&\quad-\beta_1 k_1   S_{k_1,k_2} \theta_1 -\beta_2 k_2   S_{k_1,k_2} \theta_2 \label{Sk1k2} \tag{\theequation.a}\\
\frac{\text d I_{k_1,k_2}}{\text d t} =&\quad \phantom{-} \beta_1 k_1   S_{k_1,k_2} \theta_1 +\beta_2 k_2   S_{k_1,k_2} \theta_2 - \gamma  I_{k_1,k_2}\tag{\theequation.b}\\
\frac{\text d R_{k_1,k_2}}{\text d t} =&\quad\gamma  I_{k_1,k_2}\tag{\theequation.c}
\end{alignat}
\end{subequations}

\noindent where $\beta_1$ and $\beta_2$ are the infection rates in layer $1$ and $2$, respectively.  $\theta_1$ and $\theta_2$ are analogous to the one introduced in Eq.~(\ref{theta}):
\begin{equation}
\theta_1 = \sum_{k_1} \frac{P_1(k_1)(k_1-1)}{\langle k_1\rangle}I_{k_1}  = \sum_{k_1,k_2} \frac{k_1-1}{\langle k_1\rangle}\  P(k_1,k_2)  \ I_{k_1,k_2}
\end{equation}
and the expression for $\theta_2$ is obtained by interchanging the indices.

\subsection{Basic reproduction Number in the Multiplex Model}

To obtain $\mathcal R_0$, we use the NGM as in the single-layer model. Here, we sort the $K_1\times K_2\times 3$ compartments so that the vector of variables is 
\begin{equation}\label{vector_X}
    X=(I_{1,1},I_{1,2},\ldots,I_{1,K_2}, I_{2,1},\ldots, I_{K_1,K_2}, S_{1,1},\ldots,S_{K_1,K_2},R_{1,1},\ldots,R_{K_1,K_2})
\end{equation}
To simplify notation, we flatten the two-dimensional index $(k_1,k_2)$ to a single index $i$, allowing us to represent each sub-compartment with a fixed order. We define the following mapping:

\begin{equation}
i(k_1,k_2) = k_2 + K_2\cdot(k_1-1)   
\end{equation}

The flattened index $i$ takes values from $1$ to $ \kappa\doteq K_1\times K_2$. The original indices can be recovered via the inverse mapping 

\begin{equation}
 \quad k_1(i) = \lceil i/K_2\rceil \quad\quad k_2(i)= \left\{ \begin{array}{cl}
 i\mod K_2 &: i\mod K_2\neq 0 \\
K_2 &: i\mod K_2= 0 
\end{array} \right.
\end{equation}

Using this notation, the vector of variables introduced in Eq.~(\ref{vector_X}) can be expressed as $X= (I_1,\cdots, I_\kappa, S_1,\cdots,S_\kappa,R_1,\cdots,R_\kappa)$. We can now write the system of equations (\ref{multiplex_model}) as $\dot X = \mathscr F - \mathscr V$ with
\begin{equation}
    \mathscr F_i =\left\{
    \begin{array}{cc}
        \beta_1 k_1(i)S_i\theta_1 + \beta_2 k_2(i)S_i\theta_2 & i\leq \kappa \\
         0 & \kappa< i\leq 3\kappa
    \end{array} \right.\quad\quad \mathscr V_i = \left\{
    \begin{array}{cc}
        \gamma I_i & i\leq \kappa \\
         \mathscr F_{i-\kappa} & \kappa<i\leq 2\kappa\\
         -\gamma I_{i-2\kappa }& 2\kappa< i\leq 3\kappa
    \end{array} \right.
\end{equation}

\noindent and proceed by calculating the matrix $F$:
\begin{equation}
F_{i,j}= \frac{\partial \mathscr F_i}{\partial I_j}(X=X^*) = \beta_1 k_1(i) \frac{k_1(j)-1}{\langle k_1\rangle}P(j)+\beta_2 k_2(i) \frac{k_2(j)-1}{\langle k_2\rangle}P(j)
\end{equation}
\noindent where we introduce the notation $P(j)\doteq P(k_1(j),P(k_2(j))$, and $X^*$ represents the disease free equilibrium, with $S_{i}^*=1$, $I_{i}^*=R^*_{i}=0\;\forall i$. As in the single-layered model, ${V^{-1}_{i,j}=\frac1\gamma\ \delta_{i,j}}$. In an analogous way as in the previous section, we introduce the vectors $v_1^i = k_1(i)$, ${u_1^j = [k_1(j)-1]P(j)}$, $v_2^i = k_2(i)$, and $u_2^j = [k_2(j)-1]P(j)$. For simplicity, we also define the parameters $\alpha_1$ and $\alpha_2$ as  $\alpha_i=\frac{\beta_i}{\gamma\langle k_i\rangle}$. With these notations, the matrix $M=FV^{-1}$ can be expressed as the sum of two matrices of rank one as follows:
\begin{equation}
    M = FV^{-1} =\alpha _1 \,\vec v_1 \,\vec u_1^T + \alpha _2 \,\vec v_2 \,\vec u_2^T \label{M_FV}
\end{equation}
Given that $v_1$ and $v_2$ are linearly independent, as are $u_1$ and $u_2$, it follows that $M$ has rank two, and therefore it has two non-zero eigenvalues. In \ref{appendix} we show that these are the same eigenvalues of the following matrix $J$:

\begin{equation}
 J = \begin{pmatrix}
\alpha_1 \langle \vec u_1,\vec v_1\rangle &  \alpha_2 \langle \vec u_1,\vec v_2\rangle \\ 
\alpha_1 \langle \vec u_2,\vec v_1\rangle &  \alpha_2 \langle \vec u_2,\vec v_2\rangle 
\end{pmatrix} =\begin{pmatrix}
\frac{\beta_1}{\gamma}\frac{(\langle k_1^2\rangle-\langle k_1\rangle)}{\langle k_1\rangle} & \frac{\beta_2}{\gamma} (\langle k_1\rangle-1) \\\frac{\beta_1}{\gamma} (\langle k_2\rangle-1) & \frac{\beta_2}{\gamma}\frac{\langle k_2^2\rangle-\langle k_2\rangle}{\langle k_2\rangle} 
\end{pmatrix} \label{J}
\end{equation}
If the mean degree of each layer is greater than $1$, then $J$ has two real and positive eigenvalues. This is because it guarantees that all entries are always positive and its determinant is also always positive, as shown below:
\begin{equation}
\Delta = \det J= \frac{\beta_1\beta_2}{\gamma^2\langle k_1\rangle \langle k_2\rangle} \left(\mathrm{Var}(k_1)\mathrm{Var}(k_2) + \mathrm{Var}(k_1) (\langle k_2\rangle^2-\langle k_2\rangle) + \mathrm{Var}(k_2) (\langle k_1\rangle^2-\langle k_1\rangle)\right)\label{detJ}
\end{equation}
\noindent while the trace is
\begin{equation}
\tau  =\frac{\beta_1}{\gamma} \frac{\langle k_1^2 \rangle-\langle k_1 \rangle}{\langle k_1 \rangle} +\frac{\beta_2}{\gamma} \frac{\langle k_2^2 \rangle-\langle k_2 \rangle}{\langle k_2 \rangle} \label{tau}
\end{equation}
It follows that the $\mathcal R_0$ of the model is given by
\begin{equation}
    \mathcal R_0 = \rho(J) = \frac {\tau+\sqrt{\tau^2-4\Delta}}2  \label{R0_multiplex}
\end{equation}

For networks that have low degree variance, the determinant shown in Eq.~(\ref{detJ}) is much smaller than the squared trace, $\tau^2$. Therefore, the basic reproduction number for the model can be approximated by
\begin{equation}
    \mathcal R_0 = \frac {\tau+\sqrt{\tau^2-4\Delta}}2\simeq \tau = \frac{\beta_1}{\gamma} \frac{\langle k_1^2 \rangle-\langle k_1 \rangle}{\langle k_1 \rangle} +\frac{\beta_2}{\gamma} \frac{\langle k_2^2 \rangle-\langle k_2 \rangle}{\langle k_2 \rangle}   \label{R0_tau}
\end{equation}

\noindent which constitutes a natural extension of the corresponding result for single-layer networks given by Eq.~(\ref{R0_monoplex}). When the approximation holds, the resulting $\mathcal R_0$ is simply the sum of the basic reproduction numbers associated with each layer considered separately. However, when the determinant term cannot be neglected, the exact expression becomes $\mathcal R_0=\rho(J)<\tau$, indicating that the coupling between layers is not fully captured by the additive approximation.

While $\rho \equiv \rho(J)$ provides the exact expression for $\mathcal R_0$, the quantity $\tau$ is considerably simpler to compute and admits a more transparent interpretation in terms of the epidemiological and topological parameters of the system. For instance, when comparing intervention strategies, the effects of modifying transmission rates or contact patterns within a given layer can be directly identified through the corresponding contribution to $\tau$, whereas their influence on $\rho$ is less explicit. For these reasons, in the following sections we assess the extent to which $\tau$ can serve as a practical approximation to the exact basic reproduction number.

In the following sections we present numerical results exploring the role of $\rho$ as the basic reproduction number of the system, and supporting $\tau$ as a good approximation for $\mathcal R_0$. Section~\ref{section:results_ode} includes results coming from numerically integrating the differential equations from Eq.~(\ref{multiplex_model}), while Section~\ref{section:results_agents} shows results originating from agent-based simulations.

\section{Numerical Results and Analysis} \label{section:results_ode}

Throughout this section, we present and analyze results obtained from the numerical integration of the system of Eqs.~(\ref{multiplex_model}). Our primary goal is to examine how the maximum infection peak, $\displaystyle I_\mathrm{max} = \max_t I_\mathrm{tot}(t)$, and the final epidemic size, $R_\infty = R_\mathrm{tot}(t \to \infty)$, depend on the exact expression for the basic reproduction number denoted as $\rho\doteq \rho(J)$ given in Eq.~(\ref{R0_multiplex}). Once this behavior is established, we then turn to analyze the approximation $\tau$ from Eq.~(\ref{tau}), with the aim of assessing to what extent $\tau$ can reproduce the same results as $\rho$ while offering a simpler and more transparent formulation. In particular, we test whether the epidemic threshold occurs at $\rho=1$ (or equivalently at $\tau \approx 1$), and whether the epidemic indicators can be predicted from $\tau$ alone, independently of the underlying parameters.

We focus our analysis on three types of two-layered networks. In the first type, both layers follow a Poisson degree distribution; in the second, both follow a geometric distribution; and in the third, one layer follows a Poisson distribution while the other follows an exponential one. For each network type, we examine eight scenarios: four in which $\mathcal R_0$ varies as a consequence of changes in the network topology (\textit{i.e.}, the parameters defining the degree distributions), and four in which $\mathcal R_0$ varies due to changes in the infection parameters.

\begin{figure}[ht]\centering
   \begin{subfigure}{.4\textwidth} \centering
 \includegraphics[width=\linewidth]{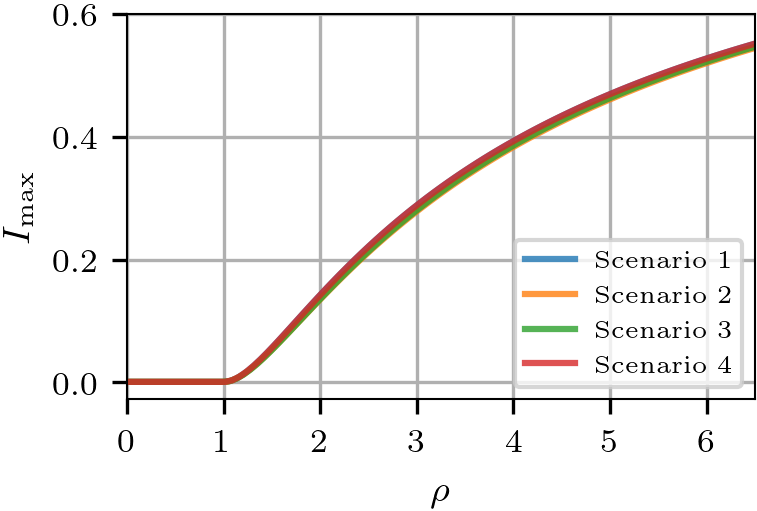}
 \end{subfigure}
   \begin{subfigure}{.4\textwidth} \centering
 \includegraphics[width=\linewidth]{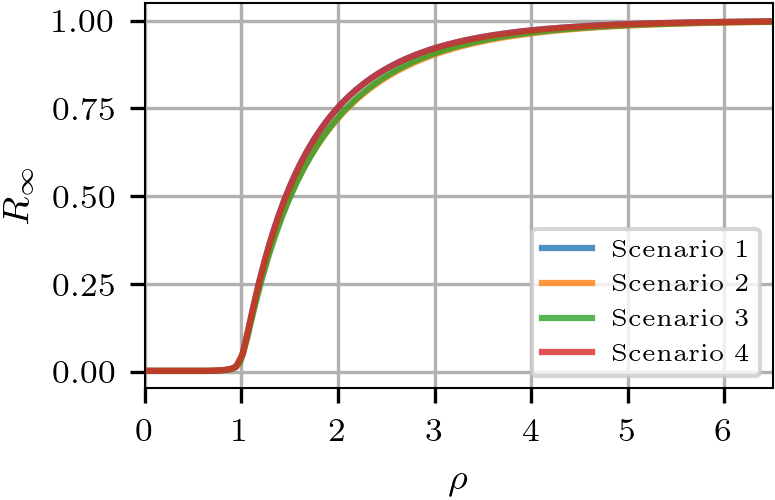}
 \end{subfigure}
    \begin{subfigure}{.4\textwidth} \centering
 \includegraphics[width=\linewidth]{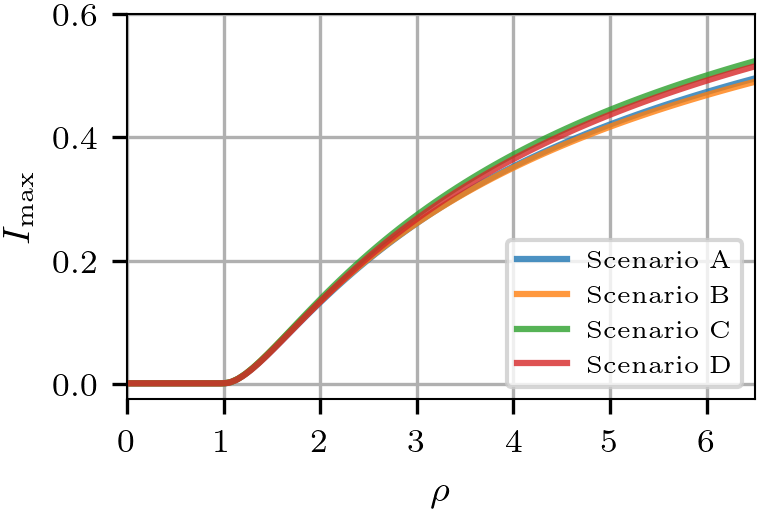}
 \end{subfigure}
   \begin{subfigure}{.4\textwidth} \centering
 \includegraphics[width=\linewidth]{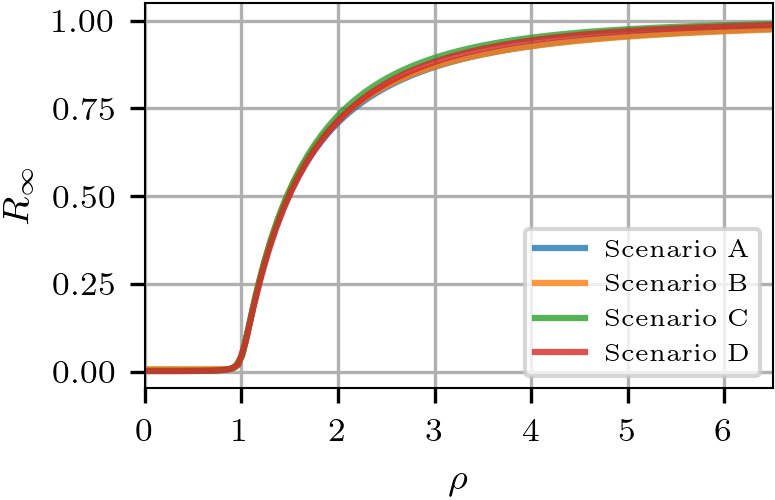}
 \end{subfigure}
 \caption{ $I_\mathrm{max}$ and $R_\infty$ as functions of the exact $\mathcal R_0$, denoted as $\rho$, given by Eq.~(\ref{R0_multiplex}). Both layers of the network have a Poisson degree distribution, $P_i(k_i)=\frac{\lambda_i^{k_i}e^{-\lambda_i}}{k_i!}$. In each considered scenario, all the parameters of the system are fixed except for one, which varies in a certain range (see Table~\ref{table_poisson} for details).  }
 \label{results_poisson_R0}
\end{figure}

Fig.~\ref{results_poisson_R0} shows the epidemic indicators as a function of $\rho $ from Eq.~(\ref{R0_multiplex}), when both layers of the network follow Poisson degree distributions with parameters $\lambda_1$ and $\lambda_2$, respectively. The maximum degree in each layer is fixed at $K_1=K_2=100$. The initial conditions are $I_{k_1,k_2}(0) = 0.001$, $S_{k_1,k_2}(0) = 1 - I_{k_1,k_2}(0)$, and $R_{k_1,k_2}(0)=0$ for all $k_1,k_2$. Details of the remaining parameters are given in Table~\ref{table_poisson}. In the scenarios labeled as 1, 2, 3 and 4, $\mathcal R_0$ is controlled by varying the network parameters ($\lambda_1$, $\lambda_2$), while the infection parameters ($\beta_1$, $\beta_2$, $\gamma$) are fixed. Conversely, in the scenarios labeled as A, B, C and D, the network parameters are kept fixed while the infection parameters are varied. The parameter values are chosen such that $\mathcal R_0$ takes values up to $6.5$.

\begin{table}[ht]\centering
\begin{tabular}{|c|c|c|c|c|c|}
\hline
Scenario & $\lambda_1$     & $\lambda_2$            & $\beta_1$              & $\beta_2$         & $1/\gamma$      \\ \hline
1    & $[0.5\, ;\,50]$ & $5$                    & $0.025$                & $0.025$           & $5$             \\ \hline
2    & $[1\, ; \,26]$  & $5$                    & $0.05$                 & $0.025$           & $5$             \\ \hline
3    & $[2\, ; \,19]$  & $\lambda_2=\lambda_1$  & $0.025$                & $0.05$            & $5$             \\ \hline
4    & $[2\, ; \,19]$  & $\lambda_2=2\lambda_1$ & $0.025$                & $0.025$           & $5$             \\ \hline
A    & $5$             & $5$                    & $[0.005\, ; \, 0.255]$ & $0.025$           & $5$             \\ \hline
B    & $4$             & $7$                    & $[0.0005\, ; \, 0.31]$ & $0.025$           & $5$             \\ \hline
C    & $4$             & $7$                    & $[0.0125\, ; \, 0.13]$ & $\beta_2=\beta_1$ & $5$             \\ \hline
D    & $5$             & $5$                    & $0.05$                 & $0.025$           & $[1\, ; \, 19]$ \\ \hline
\end{tabular}
\caption{Parameters used in each curve shown in Fig.~\ref{results_poisson_R0} and Fig.~\ref{results_poisson}. Both layers of the network have a Poisson degree distribution with mean $\lambda_1$ and $\lambda_2$ respectively.}
\label{table_poisson}
\end{table}

Fig.~\ref{results_poisson_R0} shows that for small values of $\rho$, both $I_\mathrm{max}$ and $R_\infty$ remain essentially constant until $\rho$ reaches $1$. For $\rho >1$, the epidemic is able to spread ($I_\mathrm{max}>I_0$), and both indicators increase monotonically. In a more global scale, we observe that regardless of which parameter is varied in each scenario, and independently of the values of the fixed parameters, all curves collapse onto a single trend. This demonstrates that the severity of the epidemic can be estimated solely from the value of $\rho$. In other words, both indicators depend on $\rho$ in the same way that they depend on $\mathcal R_0$ in the standard SIR model. Moreover, their behavior closely resembles the curves displayed in Fig.~\ref{ind_SIR}.

\begin{figure}[ht]\centering
  \begin{subfigure}{.49\textwidth} \centering
 \includegraphics[width=\linewidth]{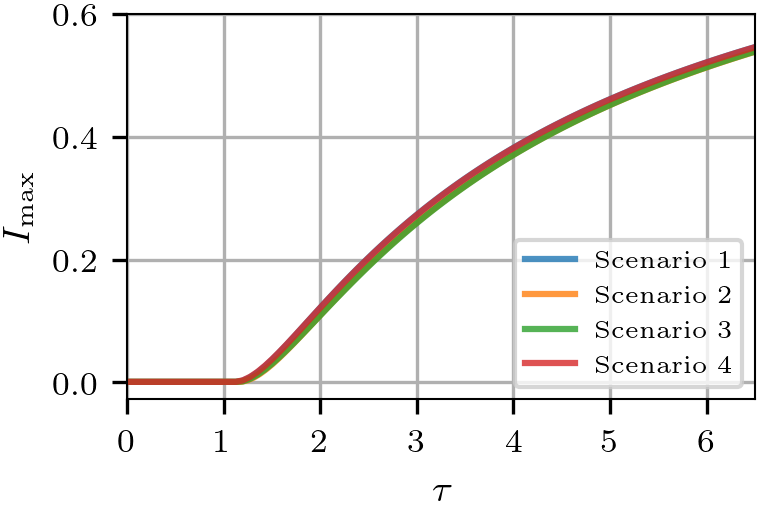}
 \end{subfigure}
   \begin{subfigure}{.49\textwidth} \centering
 \includegraphics[width=\linewidth]{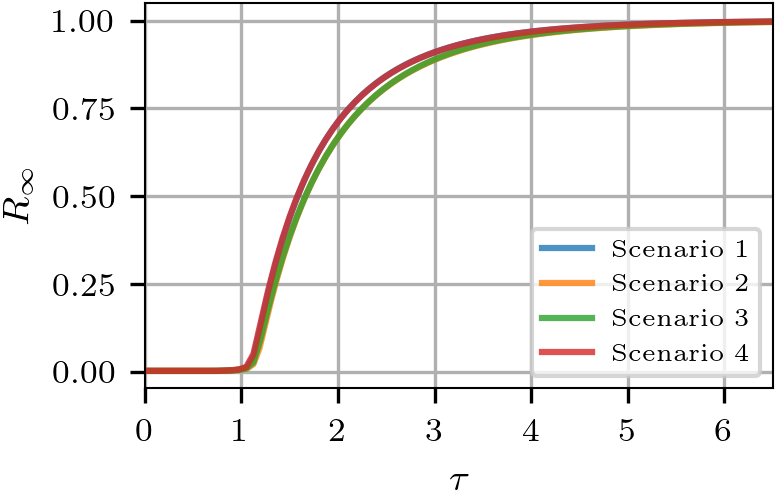}
 \end{subfigure}
 \begin{subfigure}{.49\textwidth} \centering
 \includegraphics[width=\linewidth]{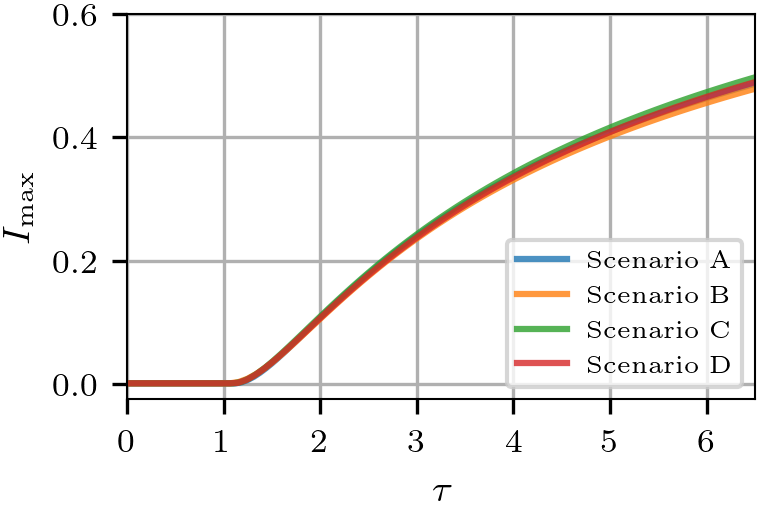}
 \end{subfigure}
   \begin{subfigure}{.49\textwidth} \centering
 \includegraphics[width=\linewidth]{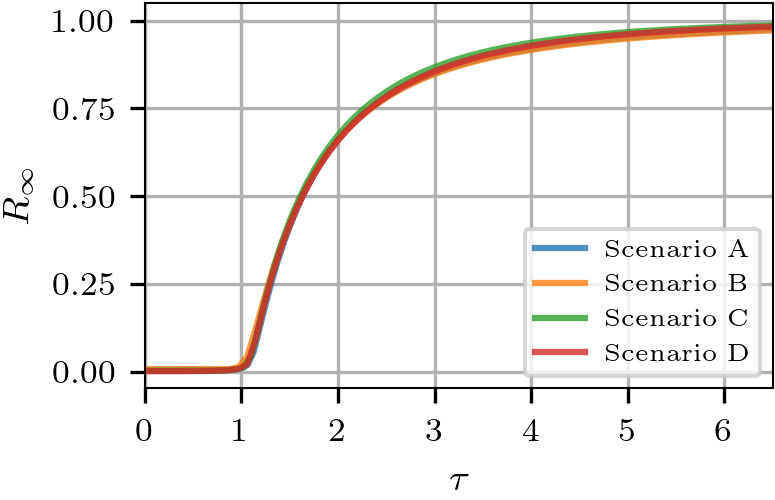}
 \end{subfigure}
 \caption{ $I_\mathrm{max}$ and $R_\infty$ as a function of $\tau = \frac{\beta_1}{\gamma}\frac{\langle k_1^2\rangle-\langle k_1\rangle}{\langle k_1\rangle} + \frac{\beta_2}{\gamma}\frac{\langle k_2^2\rangle-\langle k_2\rangle}{\langle k_2\rangle} $. Both layers of the network have a Poisson degree distribution. In each considered scenario, all the parameters of the system are fixed except for one, which varies in a certain range (see Table~\ref{table_poisson} for details).  }
 \label{results_poisson}
\end{figure}

We now turn to the approximation $\tau$ given in Eq.~(\ref{R0_tau}) to test whether it captures the same behavior as $\mathcal R_0$, while being simpler to compute and more transparent to interpret. In Fig.~\ref{results_poisson}, we present the indicators as functions of $\tau$ and compare their behavior to the exact formulation.
 
We can see in Fig.~\ref{results_poisson} that the indicators as a function of $\tau$ behave approximately in the same way as they do as a function of $\rho$: around $\tau\approx 1$, the epidemic starts spreading and then both indicators increase monotonically. The curves also collapse, meaning that the indicators also depend solely on $\tau$, regardless of the considered scenario. As in Figre~\ref{results_poisson_R0}, the indicators depend on $\tau$ in a similar way as they depend on $\mathcal R_0$ in the SIR model. A closer inspection of Fig.~\ref{results_poisson}, particularly the curves of $I_\mathrm{max}$ as a function of $\tau$, shows that the infection peak begins to grow at values of $\tau$ slightly greater than $1$. This threshold is consistent across all curves and is related to the variances of the degree distributions appearing in the determinant in Eq.~(\ref{detJ}), which in turn affects the approximation in Eq.~(\ref{R0_tau}). By contrast, Fig.~\ref{results_poisson_R0} shows that when $I_\mathrm{max}$ is plotted against $\rho$, the threshold occurs precisely at $\mathcal \rho = 1$. Nevertheless, Fig.~\ref{results_poisson} confirms that $\tau$ reproduces the epidemic behavior expected from the basic reproduction number — even though the threshold appears slightly above $\tau=1$ — while being simpler to compute and offering a more transparent interpretation than the exact expression of $\mathcal{R}_0$. For these reasons, we will use $\tau$ to present the following results, as a practical and insightful approximation to $\mathcal R_0$ in multiplex networks.

\begin{figure}[ht]\centering
\begin{subfigure}{.49\textwidth} \centering
\includegraphics[width=\linewidth]{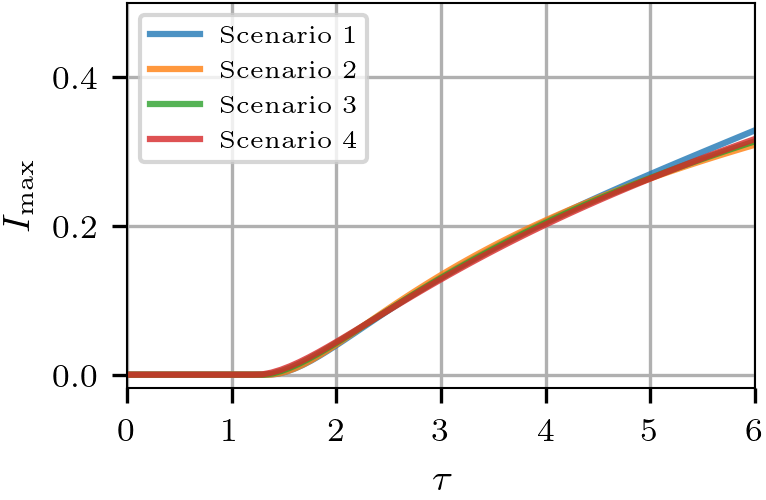}
\end{subfigure}
\begin{subfigure}{.49\textwidth} \centering
\includegraphics[width=\linewidth]{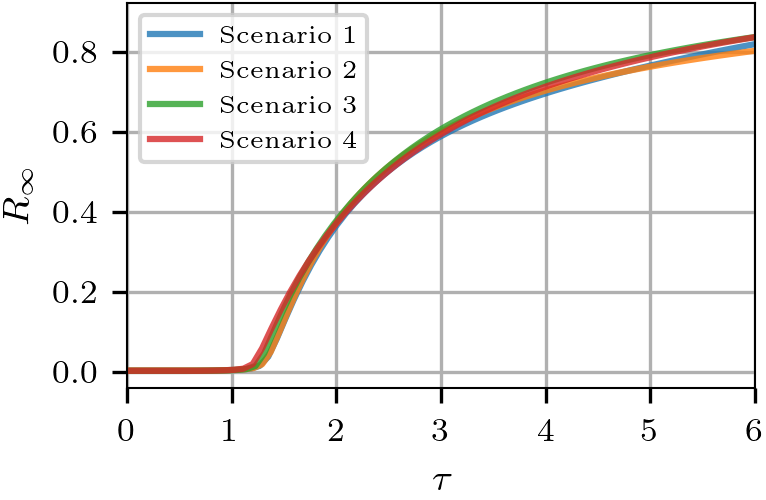}
\end{subfigure}
\begin{subfigure}{.49\textwidth} \centering
\includegraphics[width=\linewidth]{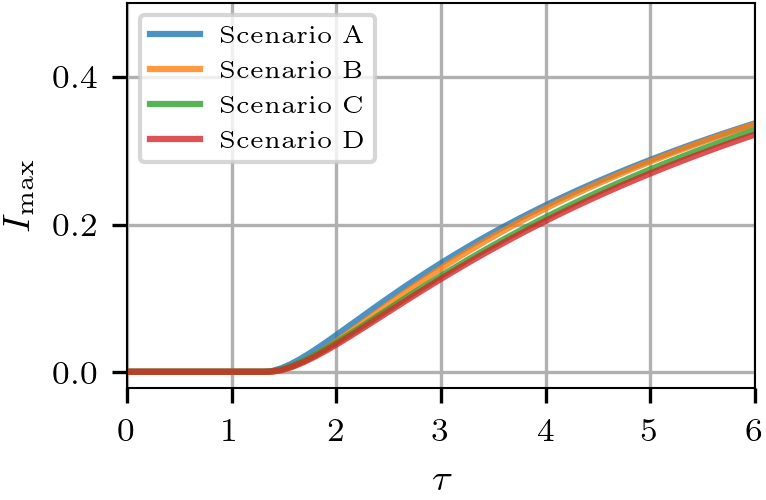}
\end{subfigure}
\begin{subfigure}{.49\textwidth} \centering
\includegraphics[width=\linewidth]{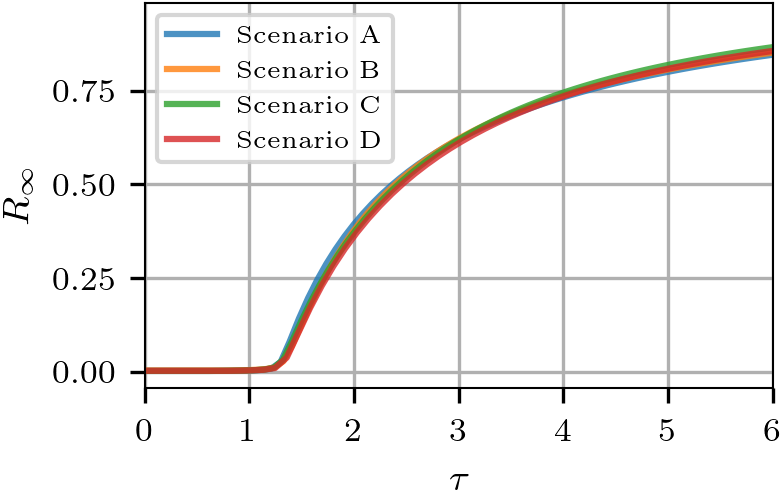}
\end{subfigure}
\caption{$I_\mathrm{max}$ and $R_\infty$ as a function of $\tau$ when both layers of the network have a geometric degree distribution, $P_i(k_i)=\mu_i^{-1}(1-\mu^{-1}_i)^{k_i-1}$. See the details of each explored scenario in Table~\ref{table_geom}. }
\label{results_geom}
\end{figure}

When both network layers follow a geometric degree distribution with mean degrees $\mu_1$ and $\mu_2$, respectively, similar results are obtained. Fig.~\ref{results_geom} presents the outcomes for eight scenarios, analogous to those discussed above, with parameter values detailed in Table~\ref{table_geom}.

\begin{table}[ht]\centering
\begin{tabular}{|c|c|c|c|c|c|}
\hline
Scenario & $\mu_1$              & $\mu_2$        & $\beta_1$              & $\beta_2$         & $1/\gamma$      \\ \hline
1    & $[1\,;\,27]$         & $4$            & $0.025$                & $0.025$           & $5$             \\ \hline
2    & $[1\, ; \,14]$       & $4$            & $0.05$                 & $0.025$           & $5$             \\ \hline
3    & $[1.75 \, ;\, 10.5]$ & $\mu_2=\mu_1$  & $0.025$                & $0.05$            & $5$             \\ \hline
4    & $[1.55\, ; \,11]$    & $\mu_2=2\mu_1$ & $0.025$                & $0.025$           & $5$             \\ \hline
A    & $3$                  & $3$            & $[0.005\, ; \, 0.35]$  & $0.025$           & $5$             \\ \hline
B    & $3$                  & $4$            & $[0.0004\, ; \, 0.32]$ & $0.025$           & $5$             \\ \hline
C    & $3$                  & $4$            & $[0.0125\, ; \, 0.15]$ & $\beta_2=\beta_1$ & $5$             \\ \hline
D    & $5$                  & $5$            & $0.025$                & $0.05$            & $[1\, ; \, 12]$ \\ \hline
\end{tabular}
\caption{Parameters used in each curve shown in Fig.~\ref{results_geom}. Both layers of the network have a geometric degree distribution with mean $\mu_1$ and $\mu_2$ respectively.}
\label{table_geom}
\end{table}

\begin{figure}[ht]\centering
\begin{subfigure}{.49\textwidth} \centering
\includegraphics[width=\linewidth]{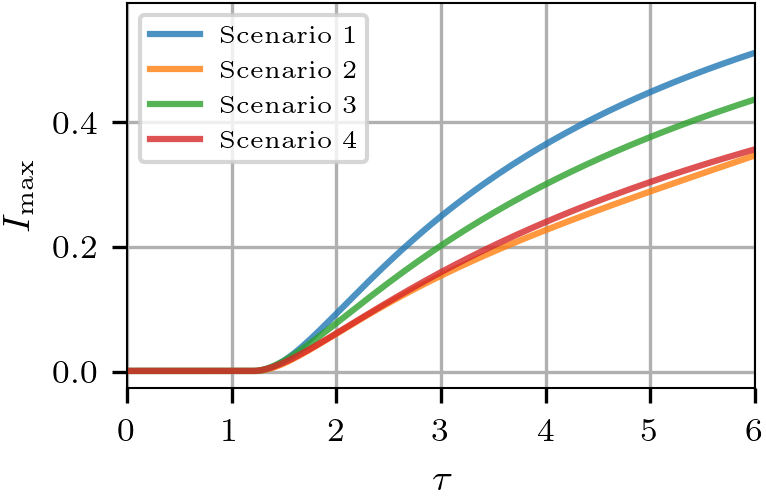}
\end{subfigure}
\begin{subfigure}{.49\textwidth} \centering
\includegraphics[width=\linewidth]{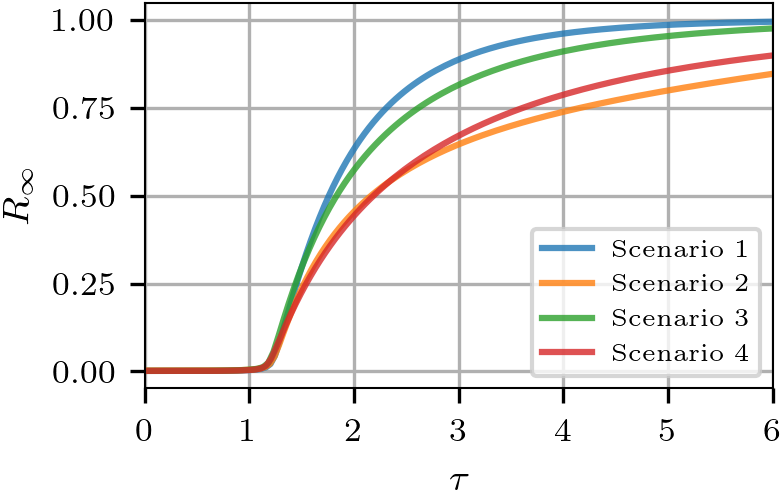}
\end{subfigure}
\begin{subfigure}{.49\textwidth} \centering
\includegraphics[width=\linewidth]{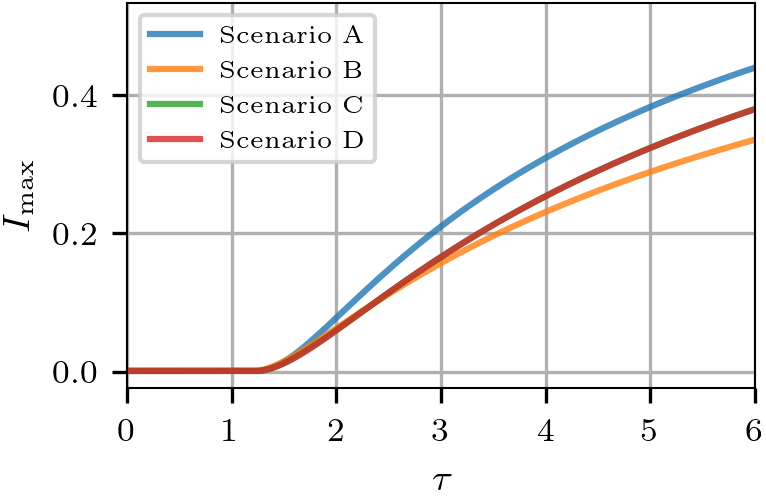}
\end{subfigure}
\begin{subfigure}{.49\textwidth} \centering
\includegraphics[width=\linewidth]{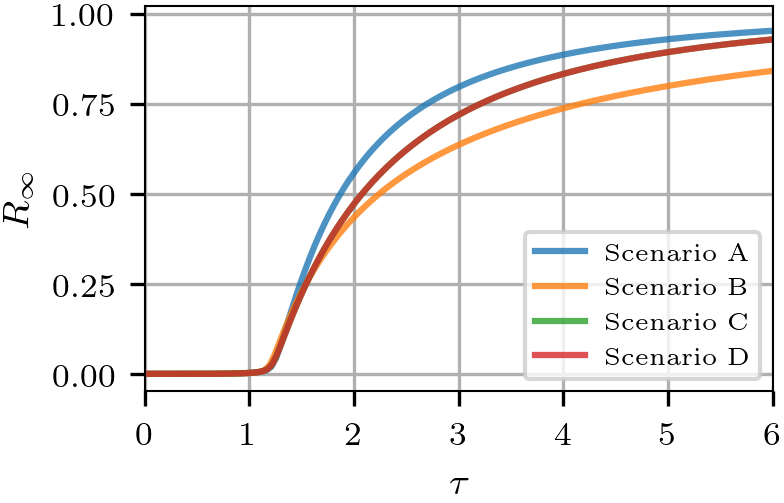}
\end{subfigure}
\caption{$I_\mathrm{max}$ and $R_\infty$ as a function of $\tau$. In all eight scenarios shown, the first layer of the network has a Poisson degree distribution with mean $\lambda$ while the second layer has a geometric one with mean $\mu$. See the details of each explored scenario in Table~\ref{table_mix}.}
\label{results_mix}
\end{figure}

However, when one layer has a Poisson degree distribution and the other has a geometric degree distribution, the results differ, as shown in Fig.~\ref{results_mix}. In this case, the curves do not collapse, meaning that $I_\mathrm{max}$ and $R_\infty$ do not depend solely $\tau$, as it also depends on the scenario being considered.

The curves shown in Fig.~\ref{results_mix} show more variance among them than the ones shown in Figs.~\ref{results_poisson} and~\ref{results_geom}. This represents a limitation on the predictability of the model, as the same value of $\tau$ can yield different values of $I_\mathrm{max}$ and $R_\infty$. However, the results do not differ more than $15\%$ for the explored scenarios, and thus, $\tau$ still provides valuable information concerning the severity of the infection. Nonetheless, for $\tau$ closer to $1$, the curves coincide, and it can be seen that $\tau\approx 1$ still acts as a threshold that divides the region where the epidemic is able to spread from the region where it does not propagate. 

\begin{table}[ht]\centering
\begin{tabular}{|c|c|c|c|c|c|}
\hline
Scenario & $\lambda$       & $\mu$            & $\beta_1$             & $\beta_2$           & $1/\gamma$      \\ \hline
1    & $[0.5\, ;\,50]$ & $3.5$            & $0.025$               & $0.025$             & $5$             \\ \hline
2    & $5$             & $[1.5 \,;\, 28]$ & $0.05$                & $0.025$             & $5$             \\ \hline
3    & $\lambda=2\mu$  & $[1.5 \,;\, 16]$  & $0.025$               & $0.05$              & $5$             \\ \hline
4    & $\lambda = \mu$ & $[1.75 \,;\, 13]$ & $0.025$               & $0.05$              & $5$             \\ \hline
A    & $4$             & $4$              & $[0.005\, ; \, 0.35]$ & $0.02$              & $5$             \\ \hline
B    & $4$             & $4$              & $0.025$               & $[0.005\,;\, 0.25]$ & $5$             \\ \hline
C    & $4$             & $4$              & $[0.01\, ; \, 0.15]$  & $\beta_2=\beta_1$   & $5$             \\ \hline
D    & $4$             & $4$              & $0.004$               & $0.04$              & $[1\, ; \, 18]$ \\ \hline
\end{tabular}
\caption{Parameters used in each curve shown in Fig.~\ref{results_mix}. The first layer has a Poisson degree distribution with mean $\lambda$ and the second one a geometric degree distribution with mean $\mu$.}
\label{table_mix}
\end{table}

The results presented in Figs.~\ref{results_poisson}, \ref{results_geom} and, to a lesser extent, in Fig.~\ref{results_mix}, support the hypothesis that $\tau$ acts as a good approximation for the basic reproduction number of the multiplex network model. Firstly, because $\tau \simeq 1$ is the epidemic threshold upon which the infection is able to grow. Secondly, for the role that $\tau$ has in estimating the severity of the epidemic outbreak, in the same way as $\mathcal R_0$ determines it in the classic SIR mean field model. 

In the following Section, we will look further into the link between the basic reproduction number and $\tau$ by means of a stochastic agent-based model.

\section{Results in agent-based multiplex network model}\label{section:results_agents}

In this Section, we introduce an agent-based multiplex network model in order to support the numerical results shown above. 

The model we present here comprises $N$ nodes within a multiplex complex network with two layers, with each node representing an individual. Each one of them can be in one of the epidemic compartments (S, I, or R), and time advances discretely in steps. 

A susceptible individual can contract the infection from each of his (her) infectious neighbors in layer $1$ at a rate $\beta_1$, and can contract the infection from each of his (her) neighbors in layer $2$ at a rate $\beta_2$. In a practical way, that means that during a timestep of duration $\delta t$ an individual denoted as $j$ will have a probability of $1 - \exp(-\beta_1\, n_1^j\, \delta t)$ of being infected through the first layer, and a probability $1 - \exp(-\beta_2\, n_2^j\, \delta t)$ of being infected through the second one, where $n_1^j$ and $n_2^j$ represent the number of infected neighbors of $j$ in the layers $1$ and $2$, respectively. Throughout this work, $\delta t$ is fixed to $1$ day.  Infectious individuals will recover after a random infection period, generated from an exponential distribution with mean time $\gamma^{-1} $. 

The first layer corresponds to an Erdös–Rényi graph~\cite{erdos_renyi} with each link between nodes having a probability $p$. The second layer is a small-world network built through the Watts-Strogatz model~\cite{watts_strogatz}, with $m$ initial neighbors and a fixed $50\%$ rewiring probability. The network layers were created with the Igraph package in Python~\cite{igraph}.

In each realization, a single individual is chosen at random to start infected, while the others $N-1$ start susceptible. For each set of parameters, we ran $100$ simulations varying the generated network layers and the initial infected individual. The simulation ends when there are no more infectious individuals. Then, knowing the degree distribution of the stochastically built networks, we can calculate their first and second moments, and we can compute the number $\tau = \frac{\beta_1}{\gamma}\frac{\langle k_1^2\rangle-\langle k_1\rangle}{\langle k_1\rangle} + \frac{\beta_2}{\gamma}\frac{\langle k_2^2\rangle-\langle k_2\rangle}{\langle k_2\rangle}$ for each simulation. 

\begin{figure}[ht]\centering
\begin{subfigure}{.49\textwidth} \centering
\includegraphics[width=\linewidth]{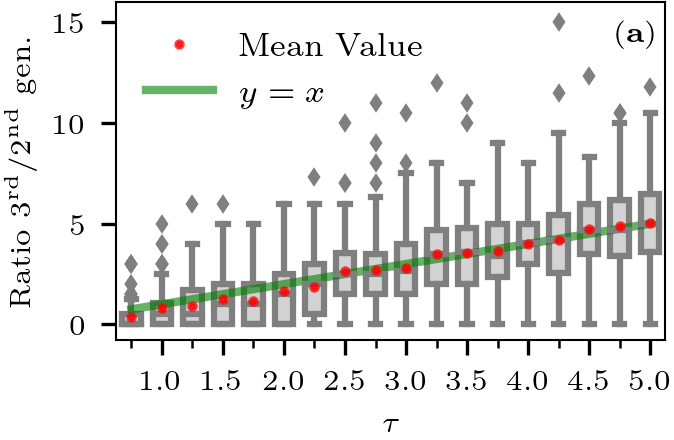}
\end{subfigure}
\begin{subfigure}{.49\textwidth} \centering
\includegraphics[width=\linewidth]{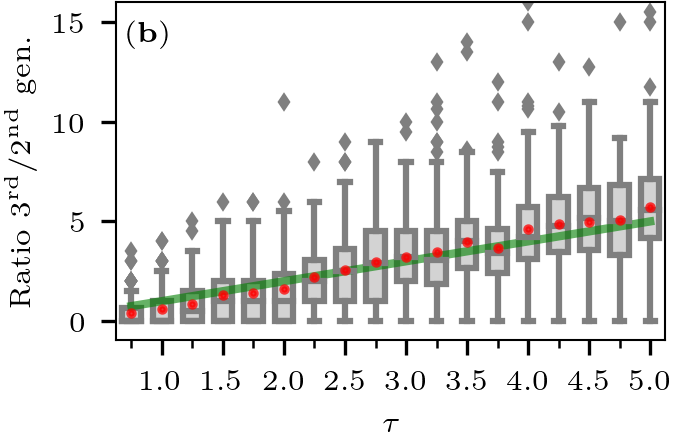}
\end{subfigure}
\begin{subfigure}{.49\textwidth} \centering
\includegraphics[width=\linewidth]{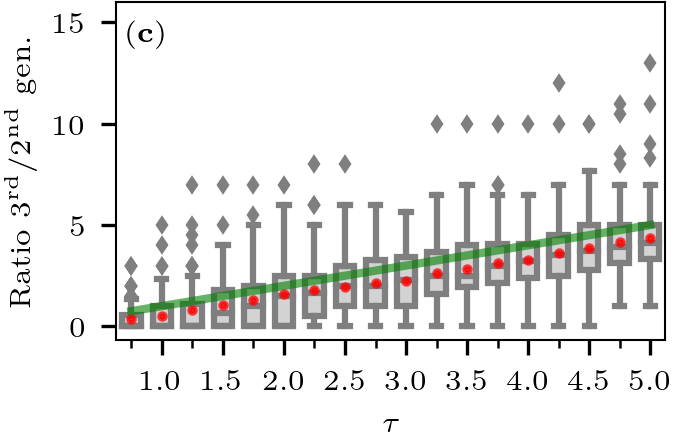}
\end{subfigure}
\caption{Box-plots of the ratio between the third and second generation of infectious individuals as a function of $\tau = \frac{\beta_1}{\gamma}\frac{\langle k_1^2\rangle-\langle k_1\rangle}{\langle k_1\rangle} + \frac{\beta_2}{\gamma}\frac{\langle k_2^2\rangle-\langle k_2\rangle}{\langle k_2\rangle}$. $\tau$ is varied by changing \textbf{(a)} the parameter $p$ in the first layer, \textbf{(b)} the parameter $m$ in the second layer, and \textbf{(c)} the infection rates $\beta_1$ and $\beta_2$. See Table~\ref{tabla_direct} for details on the values of the fixed parameters.  }
 \label{R0_direct}
\end{figure}

One commonly accepted interpretation of the basic reproduction number is the average number of secondary cases generated by a single infectious individual in an otherwise completely susceptible population. A more flexible alternative defines $\mathcal R_0$ as the average number of new infections caused by an individual when nearly all the population is still susceptible, but after several generations of transmission have occurred to reduce the impact of the random initial condition. Following the procedure in~\cite{Aparicio_2007}, we compute $\mathcal R_0$ as the ratio between the number of infections in the third and second generations: that is, the average number of new cases produced by the infectious individuals in the second generation. This ratio is shown as a function of $\tau$ in Fig.~\ref{R0_direct} for networks with $N=10^4$ nodes. In Fig.~\ref{R0_direct}.a, $\tau$ is varied by means of changing the connection probability $p$ in the Erdös–Rényi layer, while the rest of the parameters remain fixed. In Fig.~\ref{R0_direct}.b the varied parameter is $m$, the starting amount of neighbors in the Small World layer, while in Fig.~\ref{R0_direct}.c the infection rates $\beta_1$ and $\beta_2$ are varied. See Table~\ref{tabla_direct} for details on the values of the fixed parameters.

\begin{table}[ht]\centering
\begin{tabular}{|c|c|c|c|c|c|}
\hline
Fig. & $p$                     & $m$          & $\beta_1$             & $\beta_2$         & $1/\gamma$ \\ \hline
\ref{R0_direct}.a   & $[0.00011\,;\,0.00181]$ & $4$          & $0.05$                & $0.025$           & $5$        \\ \hline
\ref{R0_direct}.b    & $0.000425$              & $[1\,;\,19]$ & $0.05$                & $0.025$           & $5$        \\ \hline
\ref{R0_direct}.c   & $0.0005$                & $6$          & $[0.014\,;\,0.09325]$ & $\beta_2=\beta_1$ & $5$        \\ \hline
\end{tabular}
\caption{Values of the parameters used for Fig.~\ref{R0_direct}. }
\label{tabla_direct}
\end{table}

Fig.~\ref{R0_direct} shows that, across all box-plots, the average number of new cases coincides with the value of $\tau$. This confirms that $\tau$ effectively captures the behavior of $\mathcal R_0$, representing the average number of secondary cases generated by an infectious individual in an (almost) entirely susceptible population.

\begin{figure}[ht]\centering
  \begin{subfigure}{.49\textwidth} \centering
 \includegraphics[width=\linewidth]{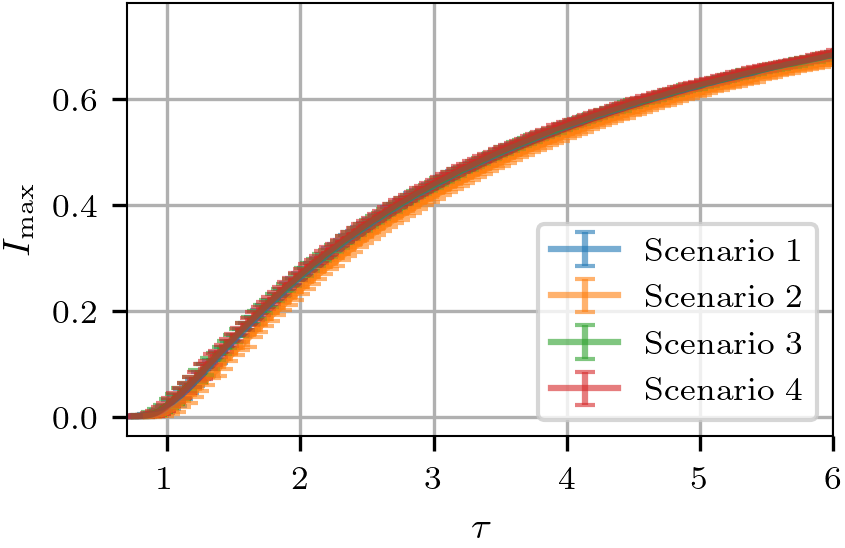}
 \end{subfigure}
   \begin{subfigure}{.49\textwidth} \centering
 \includegraphics[width=\linewidth]{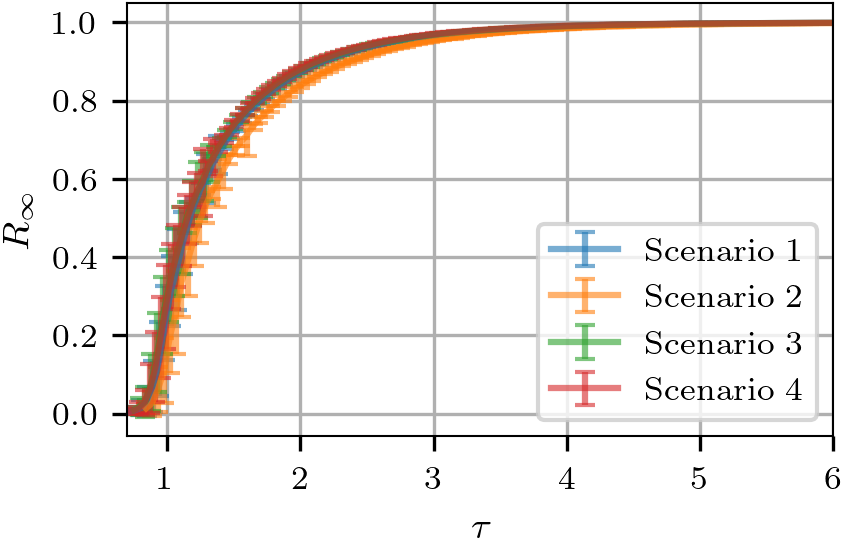}
  \end{subfigure}
   \begin{subfigure}{.49\textwidth} \centering
 \includegraphics[width=\linewidth]{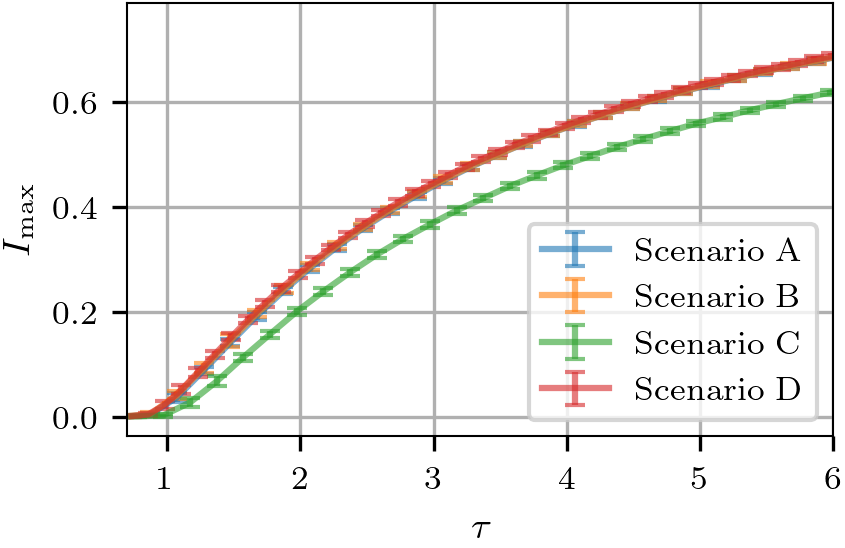}
 \end{subfigure}
   \begin{subfigure}{.49\textwidth} \centering
 \includegraphics[width=\linewidth]{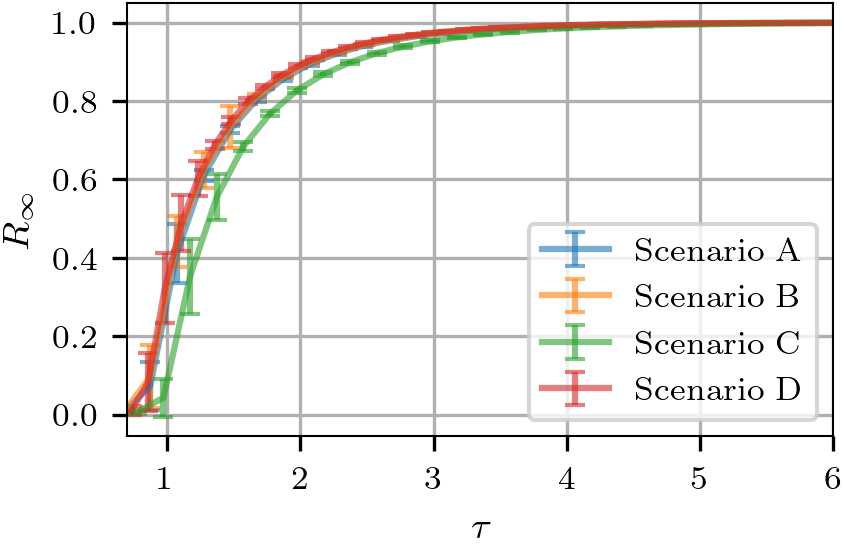}
 \end{subfigure}
   \begin{subfigure}{.49\textwidth} \centering
 \includegraphics[width=\linewidth]{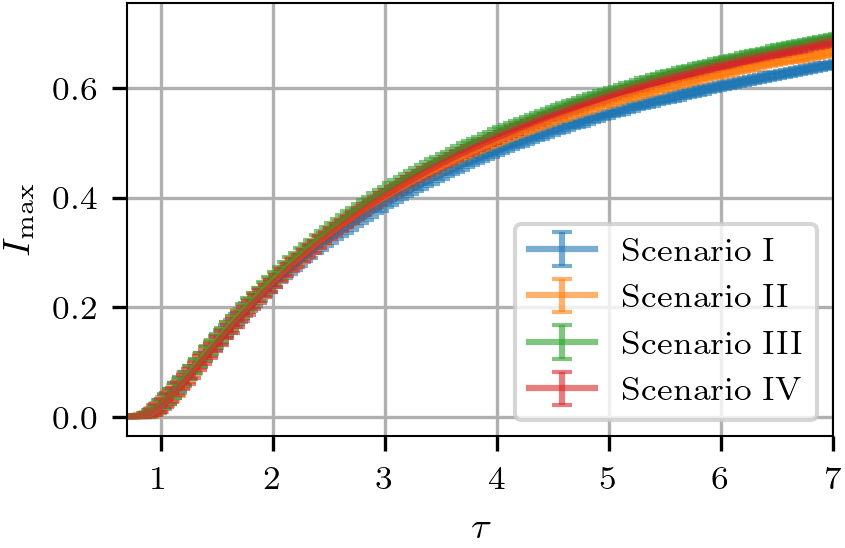}
 \end{subfigure}
   \begin{subfigure}{.49\textwidth} \centering
 \includegraphics[width=\linewidth]{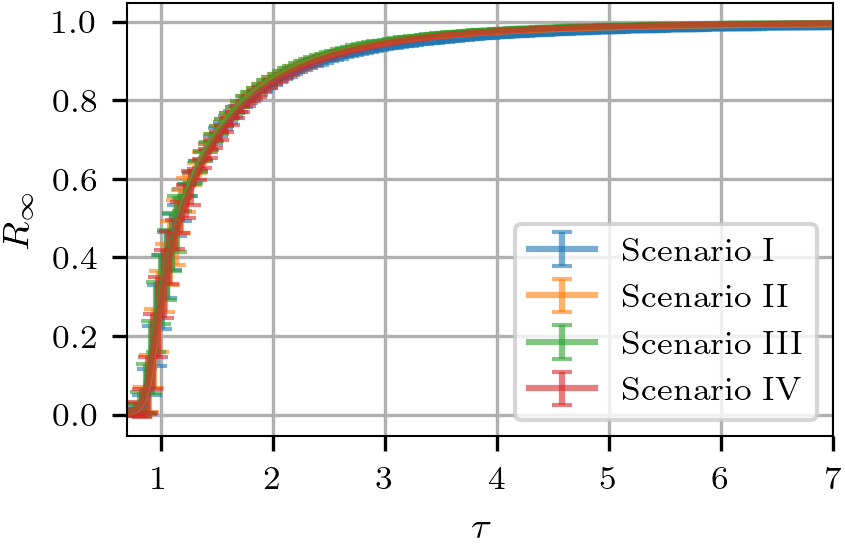}
 \end{subfigure}
 \caption{Maximum infection peak $I_\mathrm{max}$ and final epidemic size $R_\infty$ as a function of $\tau = \frac{\beta_1}{\gamma}\frac{\langle k_1^2\rangle-\langle k_1\rangle}{\langle k_1\rangle} + \frac{\beta_2}{\gamma}\frac{\langle k_2^2\rangle-\langle k_2\rangle}{\langle k_2\rangle}$ in the agent-based multiplex model. The parameters used for each set of simulations are detailed in Table~\ref{tabla_agentes}. }
 \label{indicadores_agentes}
\end{figure}

As discussed in the previous section, the basic reproduction number has a threshold functionality, as it must exceed 1 for the epidemic to spread. Moreover, it is dynamically related to key indicators such as the peak number of infected individuals, $I_\mathrm{max}$, and the final epidemic size, $R_\infty$. To illustrate the relationship between these indicators and $\tau$ when considering agent-based models, we performed simulations with $N = 10^4$ agents and $10$
individuals ($0.1\%$ of the population) randomly chosen to be initially infected. Fig.~\ref{indicadores_agentes} shows these indicators as functions of $\tau$.

\begin{table}[ht]\centering
\begin{tabular}{|c|c|c|c|c|c|}
\hline
Scenario & $p$                      & $m$           & $\beta_1$            & $\beta_2$           & $1/\gamma$ \\ \hline
1    & $[0.0002 \,;\, 0.0055]$  & $4$           & $0.025$              & $0.025$             & $5$        \\ \hline
2    & $[0.00015 \,;\,0.0055 ]$ & $4$           & $0.05$               & $0.025$             & $5$        \\ \hline
3    & $[0.00015 \,;\, 0.006]$  & $8$           & $0.0225$             & $0.015$             & $5$        \\ \hline
4    & $[0.0001\,;\,0.006]$     & $6$           & $0.02$               & $0.02$              & $6$        \\ \hline
A    & $0.0005$                 & $[1\,;\, 35]$ & $0.02$               & $0.02$              & $5$        \\ \hline
B    & $0.001$                  & $[1\,;\, 35]$ & $0.01$               & $0.02$              & $5$        \\ \hline
C    & $0.0004$                 & $[1\,;\, 35]$ & $0.01$               & $0.01$              & $10$       \\ \hline
D    & $0.001$                  & $[1\,;\, 35]$ & $0.0125$             & $0.0125$            & $5$        \\ \hline
I    & $0.0005$                 & 6             & $[0.001\,;\,0.27]$   & $0.025$             & $5$        \\ \hline
II   & $0.0005$                 & 6             & $0.025$              & $[0.001\,;\, 0.25]$ & $5$        \\ \hline
III  & $0.0005$                 & 6             & $\beta_1 = \beta_2$  & $[0.01\,;\,0.14]$   & $5$        \\ \hline
IV   & $0.0005$                 & 6             & $\beta_1 = 2\beta_2$ & $[0.0075\,;\,0.1]$  & $5$        \\ \hline
\end{tabular}
\caption{Parameters used in each curve shown in Fig.~\ref{indicadores_agentes}. The first layer is an Erdös–Rényi graph with parameter $p$, while the second layer is generated with the Watts–Strogatz model with $m$ starting neighbors and $50\%$ rewire probability. }
\label{tabla_agentes}
\end{table}
\FloatBarrier

The results presented in Fig.~\ref{indicadores_agentes} complement those shown in the previous section and reinforce the idea that, in the multiplex model, $I_\mathrm{max}$ and $R_\infty$ depend on $\tau$ in an analogous way to their dependence on $\mathcal{R}_0$ in the standard SIR model. The role as a threshold is better reflected in the behavior of $I_\mathrm{max}$, which begins to grow as $\tau$ surpasses $1$.

\section{Conclusions}
\label{section:conclusion}
In this work, we derive an analytical expression for the basic reproduction number $\mathcal R_0$ in the context of multiplex networks by means of the Next Generation Matrix method. To account for the heterogeneity and layered structure of real contact patterns, we extend the Degree-Based Mean-Field (DBMF) SIR model to multiplex networks, where individuals interact through multiple social layers, each characterized by its own degree distribution and infection rate.

Applying the Next Generation Matrix method, we obtain an exact expression for the basic reproduction number, denoted by $\rho$. Numerical integration of the multiplex DBMF equations across a variety of degree distributions, including Poisson and geometric layers, supports the theoretical predictions. In particular, the results show that $\rho=1$ marks the epidemic threshold, in agreement with the expected behavior.

We also consider an approximation denoted by $\tau$, which is simpler to compute and can be expressed as the sum of the basic reproduction numbers associated with the individual layers. Numerical results show that, although the epidemic threshold appears at values of $\tau$ slightly larger than unity, the maximum infection peak $I_\mathrm{max}$ and the final epidemic size $R_\infty$ are determined by $\tau$, independently of the specific parameter combinations used to generate it. Furthermore, both indicators increase monotonically with $\tau$ and exhibit a dependence similar to that observed in the standard SIR model as a function of $\mathcal R_0$.

Moreover, since $\tau \geq \rho$, the condition $\tau < 1$ implies $\rho < 1$. In consequence, using $\tau$ as a threshold indicator leads to a conservative estimation of the epidemic risk. While $\tau$ does not exactly reproduce the epidemic threshold, its simple analytical form and transparent interpretation in terms of the epidemiological and topological parameters of the system make it a useful approximation in practical applications.  

Agent-based simulations further support the use of $\tau$ as a meaningful epidemiological indicator. The simulations show that $\tau$ closely corresponds to the average number of secondary infections generated during the early stages of an outbreak, consistent with the epidemiological interpretation of $\mathcal R_0$. Moreover, the epidemic indicators obtained from the agent-based model exhibit functional relationships with $\tau$ that closely resemble those predicted by the standard SIR model when expressed in terms of $\mathcal R_0$.

Taken together, these results show that $\rho$ provides a rigorous extension of the basic reproduction number to multiplex networks, while $\tau$ offers a simple and interpretable approximation that preserves its main epidemiological features. The framework developed here highlights the influence of multiplex network topology on epidemic dynamics and provides tools for understanding disease transmission in populations with layered contact structures.

Finally, this study opens several directions for future research, including extending the multiplex framework to more than two layers,  vaccination, and applying these methods to empirical multiplex networks during real epidemics. In addition, while the assumption of statistically independent layers allows us to isolate the effects of multiplex structure and derive analytical results, real systems may exhibit interlayer degree correlations that could influence epidemic dynamics. Extending the framework to account for such correlations constitutes a natural direction for future research.

\section*{Acknowledgments}
The authors acknowledge financial support from CONICET and CNEA. All public Institutions from Argentina.
\section*{Compliance with Ethical Standards}
There are no conflicts of interest associated with this publication and there has been no significant financial support for this work that could have influenced its outcome. This article does not contain any studies with human participants or animals performed by any of the authors. All results are purely mathematical in nature, either analytical or numerical. There are no associated data sets from other sources.

\appendix
\section{ }\label{appendix}
In this appendix we proof that the matrix $M$ of Eq.~(\ref{M_FV}) has the same non-zero eigenvalues as the matrix $J$ from Eq.~(\ref{J}). As usual, the eigenvalues $\lambda$ and eigenvectors $\vec w$ satisfy the relation 

\begin{equation}
M\vec w = \alpha _1 \,\vec v_1 \,\vec u_1^T \,\vec w  + \alpha _2 \,\vec v_2 \,\vec u_2^T \,\vec w =\lambda \vec w 
\end{equation}
By calling the inner products $c_i \doteq \vec u_i^T \vec w = \langle \vec u_i,\vec w\rangle$, the eigenvectors can be expressed as
\begin{equation}
\vec w = \frac{\alpha_1 c_1}\lambda \,\vec v_1 + \frac{\alpha_2 c_2}\lambda\, \vec v_2 \label{ec_w}
\end{equation}

By inserting $\vec w$ from Eq.~\ref{ec_w} into the definitions of $c_1$ and $c_2$, we get
\begin{alignat}1
c_1 = \langle \vec u_1,\vec w\rangle = \left\langle \vec u_1,\frac{\alpha_1 c_1}\lambda \vec v_1 + \frac{\alpha_2 c_2}\lambda \vec v_2\right\rangle = \frac{\alpha_1 c_1}\lambda \langle \vec u_1,\vec v_1\rangle +  \frac{\alpha_2 c_2}\lambda \langle \vec u_1,\vec v_2\rangle \\ 
c_2 = \langle \vec u_2,\vec w\rangle = \left\langle \vec u_2,\frac{\alpha_1 c_1}\lambda \vec v_1 + \frac{\alpha_2 c_2}\lambda \vec v_2\right\rangle = \frac{\alpha_1 c_1}\lambda \langle \vec u_2,\vec v_1\rangle +  \frac{\alpha_2 c_2}\lambda \langle \vec u_2,\vec v_2\rangle 
\end{alignat}
This can be expressed in a more condensed way as
\begin{equation}
J \begin{pmatrix}
c_1\\c_2
\end{pmatrix} =
\begin{pmatrix}
\alpha_1 \langle \vec u_1,\vec v_1\rangle &  \alpha_2 \langle \vec u_1,\vec v_2\rangle \\ 
\alpha_1 \langle \vec u_2,\vec v_1\rangle &  \alpha_2 \langle \vec u_2,\vec v_2\rangle 
\end{pmatrix} \begin{pmatrix}
c_1\\c_2
\end{pmatrix} = \lambda \begin{pmatrix}
c_1\\c_2
\end{pmatrix} 
\end{equation}
\noindent which implies that the eigenvalues $\lambda$ of $M$ are the same as those from the matrix $J$. The inner products appearing in the components of $J$ are:
\begin{equation}
\langle \vec u_1,\vec v_1\rangle = \sum_{j=1}^\kappa k_1(j)\cdot(k_1(j)-1)P(j) = \langle k_1^2\rangle-\langle k_1\rangle
\end{equation}

\begin{equation}
\langle \vec u_1,\vec v_2\rangle = \sum_{j=1}^\kappa (k_1(j)-1)k_2(j) P(j) = (\langle k_1\rangle-1)\langle k_2\rangle
\end{equation}

In analogous way, $\langle \vec u_2,\vec v_1\rangle = (\langle k_2\rangle-1)\langle k_1\rangle$ and $\langle \vec u_2,\vec v_2\rangle = \langle k_2^2\rangle-\langle k_2\rangle$. Finally, recalling that $\alpha_i=\frac{\beta_i}{\gamma\langle k_i\rangle}$, the resulting matrix is 

\begin{equation}
 J = \begin{pmatrix}
\frac{\beta_1}{\gamma}\frac{\langle k_1^2\rangle-\langle k_1\rangle}{\langle k_1\rangle} & \frac{\beta_2}{\gamma} (\langle k_1\rangle-1) \\\frac{\beta_1}{\gamma} (\langle k_2\rangle-1) & \frac{\beta_2}{\gamma}\frac{\langle k_2^2\rangle-\langle k_2\rangle}{\langle k_2\rangle} 
\end{pmatrix} 
\end{equation}

\bibliographystyle{elsarticle-num}

\begin{thebibliography}{10}
\expandafter\ifx\csname url\endcsname\relax
  \def\url#1{\texttt{#1}}\fi
\expandafter\ifx\csname urlprefix\endcsname\relax\def\urlprefix{URL }\fi
\expandafter\ifx\csname href\endcsname\relax
  \def\href#1#2{#2} \def\path#1{#1}\fi

\bibitem{kerm1}
W.~O. Kermack, A.~G. McKendrick, {A Contribution to the Mathematical Theory of Epidemics}, Proc. R. Soc. A 115 (1927) 700.
\newblock \href {https://doi.org/10.1098/rspa.1927.0118} {\path{doi:10.1098/rspa.1927.0118}}.

\bibitem{murray}
J.~D. Murray (Ed.), {Mathematical Biology}, Vol.~17, Springer New York, New York, NY, 2002.
\newblock \href {https://doi.org/10.1007/b98868} {\path{doi:10.1007/b98868}}.

\bibitem{R0_heterogeneo}
R.~M. May, A.~L. Lloyd, {Infection dynamics on scale-free networks}, Phys. Rev. E 64 (2001) 66112.
\newblock \href {https://doi.org/10.1103/PhysRevE.64.066112} {\path{doi:10.1103/PhysRevE.64.066112}}.

\bibitem{may_HIV}
R.~M. May, S.~Gupta, A.~R. McLean, {Infectious disease dynamics: what characterizes a successful invader?}, Philosophical Transactions of the Royal Society of London. Series B: Biological Sciences 356~(1410) (2001) 901--910.
\newblock \href {https://doi.org/10.1098/rstb.2001.0866} {\path{doi:10.1098/rstb.2001.0866}}.

\bibitem{next_gen}
P.~van~den Driessche, J.~Watmough, {Reproduction numbers and sub-threshold endemic equilibria for compartmental models of disease transmission}, Mathematical Biosciences 180~(1-2) (2002) 29--48.
\newblock \href {https://doi.org/10.1016/S0025-5564(02)00108-6} {\path{doi:10.1016/S0025-5564(02)00108-6}}.

\bibitem{review_networks}
R.~Pastor-Satorras, C.~Castellano, P.~Van~Mieghem, A.~Vespignani, {Epidemic processes in complex networks}, Rev. Mod. Phys. 87~(3) (2015) 925--979.
\newblock \href {https://doi.org/10.1103/RevModPhys.87.925} {\path{doi:10.1103/RevModPhys.87.925}}.

\bibitem{kuperman_2013}
M.~N. Kuperman, {Invited review: Epidemics on social networks}, Papers in Physics 5 (2013) 050003.
\newblock \href {https://doi.org/10.4279/pip.050003} {\path{doi:10.4279/pip.050003}}.

\bibitem{networks_and_epidemic}
M.~J. Keeling, K.~T. Eames, {Networks and epidemic models}, Journal of The Royal Society Interface 2~(4) (2005) 295--307.
\newblock \href {https://doi.org/10.1098/rsif.2005.0051} {\path{doi:10.1098/rsif.2005.0051}}.

\bibitem{kuperman}
M.~Kuperman, G.~Abramson, {Small World Effect in an Epidemiological Model}, Physical Review Letters 86 (2001) 2909--2912.
\newblock \href {https://doi.org/10.1103/PhysRevLett.86.2909} {\path{doi:10.1103/PhysRevLett.86.2909}}.

\bibitem{zanette}
D.~Zanette, S.~Risau, {Infection Spreading in a Population with Evolving Contacts}, Journal of Biological Physics 34 (2008) 135--148.
\newblock \href {https://doi.org/10.1007/s10867-008-9060-9} {\path{doi:10.1007/s10867-008-9060-9}}.

\bibitem{zanette2}
D.~H. Zanette, M.~Kuperman, {Effects of immunization in small-world epidemics}, Physica A: Statistical Mechanics and its Applications 309~(3) (2002) 445--452.

\bibitem{kara}
A.~Karaivanov, {A social network model of COVID-19}, PLOS ONE 15 (2020) 1--33.
\newblock \href {https://doi.org/10.1371/journal.pone.0240878} {\path{doi:10.1371/journal.pone.0240878}}.

\bibitem{multi_agent}
F.~A. Salem, U.~F. Moreno, {A Multi-Agent-Based Simulation Model for the Spreading of Diseases Through Social Interactions During Pandemics}, Journal of Control, Automation and Electrical Systems 33~(4) (2022) 1161--1176.
\newblock \href {https://doi.org/10.1007/s40313-022-00920-3} {\path{doi:10.1007/s40313-022-00920-3}}.

\bibitem{benitez}
{H. E. Ben{\'{i}}tez}, {F. E. Cornes}, {C. O. Dorso}, {G.A. Frank}, {Disease Spreading through Complex Small World Networks}, European Society of Medicine (2024).
\newblock \href {https://doi.org/10.18103/mra.v12i9.5706} {\path{doi:10.18103/mra.v12i9.5706}}.

\bibitem{moreno2002}
Y.~Moreno, R.~Pastor-Satorras, A.~Vespignani, {Epidemic outbreaks in complex heterogeneous networks}, The European Physical Journal B - Condensed Matter and Complex Systems (2002).
\newblock \href {https://doi.org/10.1140/epjb/e20020122} {\path{doi:10.1140/epjb/e20020122}}.

\bibitem{dynamical_patterns}
M.~Barth{\'{e}}lemy, A.~Barrat, R.~Pastor-Satorras, A.~Vespignani, {Dynamical patterns of epidemic outbreaks in complex heterogeneous networks}, Journal of Theoretical Biology 235~(2) (2005) 275--288.
\newblock \href {https://doi.org/10.1016/j.jtbi.2005.01.011} {\path{doi:10.1016/j.jtbi.2005.01.011}}.

\bibitem{inmunization_pastor_satorras}
R.~Pastor-Satorras, A.~Vespignani, {Immunization of complex networks}, Phys. Rev. E 65~(3) (2002) 36104.

\bibitem{velocity_spread}
M.~Barth{\'{e}}lemy, A.~Barrat, R.~Pastor-Satorras, A.~Vespignani, {Velocity and Hierarchical Spread of Epidemic Outbreaks in Scale-Free Networks}, Phys. Rev. Lett. 92, 178701 (2004).
\newblock \href {https://doi.org/10.1103/PhysRevLett.92.178701} {\path{doi:10.1103/PhysRevLett.92.178701}}.

\bibitem{rozan_2023}
E.~Rozan, S.~Bouzat, M.~Kuperman, {Testing lockdown measures in epidemic outbreaks through mean-field models considering the social structure}, Physica A: Statistical Mechanics and its Applications 632 (2023) 129330.
\newblock \href {https://doi.org/10.1016/j.physa.2023.129330} {\path{doi:10.1016/j.physa.2023.129330}}.

\bibitem{singapore}
N.~N. Chung, L.~Y. Chew, {Modelling Singapore COVID-19 pandemic with a SEIR multiplex network model}, Scientific Reports 11~(1) (2021) 10122.
\newblock \href {https://doi.org/10.1038/s41598-021-89515-7} {\path{doi:10.1038/s41598-021-89515-7}}.

\bibitem{liuq}
Q.-H. Liu, M.~Ajelli, A.~Aleta, S.~Merler, Y.~Moreno, A.~Vespignani, {Measurability of the epidemic reproduction number in data-driven contact networks}, Proceedings of the National Academy of Sciences 115~(50) (2018) 12680--12685.
\newblock \href {https://doi.org/10.1073/pnas.1811115115} {\path{doi:10.1073/pnas.1811115115}}.

\bibitem{overlapped_multiplex_networks}
C.~Buono, L.~G. Alvarez-Zuzek, P.~A. Macri, L.~A. Braunstein, {Epidemics in Partially Overlapped Multiplex Networks}, PLoS ONE 9~(3) (2014) e92200.
\newblock \href {https://doi.org/10.1371/journal.pone.0092200} {\path{doi:10.1371/journal.pone.0092200}}.

\bibitem{brauer}
F.~Brauer, P.~van~den Driessche, J.~Wu (Eds.), {Mathematical Epidemiology}, Vol. 1945, Springer Berlin Heidelberg, Berlin, Heidelberg, 2008.
\newblock \href {https://doi.org/10.1007/978-3-540-78911-6} {\path{doi:10.1007/978-3-540-78911-6}}.

\bibitem{erdos_renyi}
P.~Erd{\H{o}}s, A.~R{\'{e}}nyi, {On random graphs. I.}, Publicationes Mathematicae Debrecen (1959) 290--297\href {https://doi.org/10.5486/PMD.1959.6.3-4.12} {\path{doi:10.5486/PMD.1959.6.3-4.12}}.

\bibitem{watts_strogatz}
D.~J. Watts, S.~H. Strogatz, {Collective dynamics of ‘small-world’ networks}, Nature 393~(6684) (1998) 440--442.
\newblock \href {https://doi.org/10.1038/30918} {\path{doi:10.1038/30918}}.

\bibitem{igraph}
C.~G{\'{a}}bor, N.~Tam{\'{a}}s, {The igraph software package for complex network research}, InterJournal Complex Systems (2006) 1695\href {https://doi.org/10.5281/zenodo.3630268} {\path{doi:10.5281/zenodo.3630268}}.

\bibitem{Aparicio_2007}
J.~P. Aparicio, M.~Pascual, {Building epidemiological models from R0 : an implicit treatment of transmission in networks}, Proceedings of the Royal Society B: Biological Sciences 274~(1609) (2007) 505--512.
\newblock \href {https://doi.org/10.1098/rspb.2006.0057} {\path{doi:10.1098/rspb.2006.0057}}.

\end{thebibliography}

\end{document}